\documentclass[
reprint,
superscriptaddress,
amsmath,amssymb,
pre,noeprint
]{revtex4-1}

\usepackage{color}
\usepackage{mathtools}
\usepackage{hyperref}
\usepackage{amsmath,amssymb,amsfonts}
\usepackage{graphicx}
\usepackage{dcolumn}
\usepackage{bm}
\usepackage{float}
\usepackage{epstopdf}
\usepackage{textcomp}
\usepackage{verbatim}

\def\be{\begin{equation}}
\def\ee{\end{equation}}
\def\bea{\begin{eqnarray}}
\def\eea{\end{eqnarray}}
\def\bsplit{\begin{split}}
	\def\esplit{\end{split}}

\def\p{\partial} 
\def\nn{\nonumber}
\def\f{\frac}

\def\bs{\boldsymbol}

\def\l{\left(}
\def\r{\right)}
\def\ls{\left[}
\def\rs{\right]}

\def\la{\langle}
\def\ra{\rangle}
\def\lla{\left\langle}
\def\rla{\right\rangle}
\def\bs{\boldsymbol}

\def\mr{\mathrm}
\def\refn{Eq.\,\ref}
\def\xp{x_{\perp}}

\newcommand{\ncbs}{\affiliation{Simons Centre for the Study of Living Machines, National Centre for Biological Sciences (TIFR), Bangalore 560065, India}}
\newcommand{\mbi}{\affiliation{Mechanobiology Institute and Department of Biological Sciences, National University of Singapore, 117411 Singapore}}
\newcommand{\curie}{\affiliation{Laboratoire Physico Chimie Curie, Institut Curie, PSL Research University, CNRS UMR168, 75005 Paris, France}}

\begin{document}
	
	\graphicspath{{figures/}}

	\title{Breakdown of effective temperature, power law interactions and self-propulsion in a momentum conserving active fluid
	}
	
	\author{Amit Singh Vishen}
	\thanks{Present address: Laboratoire Physico Chimie Curie, Institut Curie, PSL Research University, CNRS UMR168, 75005 Paris, France}
	\ncbs
	
	\author{  Jacques Prost}
	\email{jacques. prost@curie.fr}
	\mbi
	\curie
	
	\author{Madan Rao}%
	\email{madan@ncbs.res.in}
	\ncbs

	\date{\today}%
	
	\begin{abstract}
	The simplest extensions of single particle dynamics
	in a momentum conserving active  fluid - an active suspension of two colloidal particles or a single particle confined by a wall -  
	exhibit strong departures from 
	Boltzmann behavior, resulting in either a breakdown of an effective temperature description or a steady state with nonzero entropy production rate. This is a consequence of 
	hydrodynamic interactions that introduce multiplicative noise in the
	stochastic description of particle positions. 
	This results in fluctuation-induced 
interactions that depend on distance as a power law.
	We find that
	the dynamics of activated colloids in a passive fluid,
	with stochastic forcing localized on the particle, is different from that of passive colloids in an active fluctuating fluid. 
	\end{abstract}

	\maketitle
	
\section{Introduction}

	Fluctuations of a dilute active suspension (e.g., bacterial bath,~\cite{Wu2000,Lau2003}) have often been described in terms of an equilibrium system with 
	a (large) effective temperature~\cite{Palacci2010, Takatori2015},
	with theoretical rationalisations provided by studies of the dynamics of a single particle in an active fluctuating
	fluid or a single active particle embedded in a passive fluid \cite{Takatori2015, Burkholder2017}. 
	
	However, as we find here, even the simplest extensions - the stochastic dynamics of 
	{\it two} particles embedded in an unbounded 
	isotropic active fluctuating gel or a particle in an isotropic active fluctuating gel bound within confining walls - do not allow for an effective temperature description, since the corresponding steady state probability
	distribution shows strong departures from the Boltzmann form. This is a consequence of a drift that arises from hydrodynamic interactions, that introduces a multiplicative noise in the
	stochastic description of particle positions. The form of the drift can only be decided {\it after solving the full hydrodynamics problem}.
	This has important implications for current discussions on active contributions to pressure, osmotic pressure and surface tension in momentum conserving active fluids \cite{Takatori2015, Marconi2015, Rodenburg2017}.
	
	Indeed, deviations from Boltzmann behaviour and consequent  
	breakdown of an effective temperature description
	have been systematically analysed in a {\it dry} system of active Ornstein-Uhlenbeck particles (AOUP)~\cite{Fodor2016}, where momentum is not conserved.
	Within a systematic perturbation expansion in the active noise correlation time 
	$\tau_n$, the nonequilibrium nature of the steady state distribution first shows up at order $\tau_n$ (characterised by non-Boltzmann
	probability distribution but zero entropy dissipation), while the full nonequilibrium aspect with nonzero entropy production shows up at order $\tau_n^{3/2}$~\cite{Fodor2016}.
Moreover, there have been several studies on the form of the effective interaction between passive particles (both fixed and mobile) embedded in a bath of dry active particles. For instance, passive colloids in a suspension of self-propelled particles interact through a non-equilibrium analog of depletion forces \cite{Angelani2011, Ray2014, Ni2015}. The range of these interactions depends on the shape of tracers - being short-range between passive spheres \cite{Angelani2011}, and long-range between two parallel walls 
 \cite{Ni2015}. In addition, passive spheres can have large interactions due to density fluctuations of dry active particles \cite{Baek2017}. 
 
In this paper, we ask, what is the nature of effective interactions and departures from Boltzmann distribution, in momentum conserving active systems?
		We find that the simplest extensions of single-particle dynamics, viz., that of colloid particle-wall and colloid particle-particle interactions embedded in a three dimensional active 
	fluctuating gel, exhibits a clear non-Boltzmann steady state distribution, 
	characterized by an effective attractive potential 
    ($\propto 1/r $, for particle-wall separated by $r$ and $\propto 1/r^4$, for particle-particle). 
    This is a consequence of the active (fluctuation-dissipation relation violating) 
    fluctuations and
	hydrodynamic interactions that introduce a multiplicative noise. 
	 A dimer of unequal sized spherical particles embedded in an active fluctuating gel self-propels with a velocity proportional to the fluctuation amplitude.  
	         Interestingly, for this  momentum conserving active system, both the breakdown of the effective temperature description
		and finite entropy production rate, {\it appear even in the limit $\tau_n \to 0$}.
	We next study the statistics of fluctuations of activated particles in a passive fluid, i.e., particles directly subject to a stochastic driving force. We find 
	that the dynamics of activated particles in a passive medium
	 is {\it not} the same as the dynamics of passive particles in an active
		medium - 
		for instance, the effective particle-particle interaction is repulsive and long ranged ($\propto 1/r^2$). 		
		We proceed to demonstrate these results below. 
			
	Consider an incompressible, isotropic, actively fluctuating viscoelastic gel, described by a local stress,
	\be
	\l 1 + \tau_v \p_t \r \sigma_{ij} = - p + \eta \l \p_i v_j + \p_j v_i \r + \sigma^n_{ij},
	\label{eq:stress1}
	\ee
	where $i,j \in (x,y,z)$, $\tau_v$ is a Maxwell time, $\eta$ is the viscosity, $p$ is the pressure which includes, a priori, the isotropic component of the mean and fluctuating active stress, and $ \sigma^n_{ij}$ is the fluctuating component of the active deviatoric stress,
	with zero mean and correlation~\cite{Lau2009, Basu2008}
	\bea
	& \la \sigma^n_{ij}({\bf r},t) & \sigma^n_{kl}({\bf r'}, t') \ra  =  \nn \\
	& & 2\pi  \delta({\bf r - r'}) \Delta(t - t') \left[ \delta_{ik} \delta_{jl}  +  \delta_{il} \delta_{jk}    - \f{2}{3}  \delta_{ij} \delta_{kl}  \right] , \quad 
	\label{eq:noise_corr}
	\eea
	with $\Delta(t-t') = \Lambda \tau_n^{-1} e^{-\vert t - t'\vert/\tau_n}$. For 
	simplicity, we have taken the variance of the anisotropic stress fluctuation to be a scalar $\Delta(t - t')$. In general, $\Delta_{ijkl}(t-t')$ is a fourth rank tensor, 
		which can arise from fluctuations of the nematic order parameter~\cite{Lau2009}.
Since the temporal correlations of the noise are unrelated to the drag ($\tau_v \neq \tau_n$), this system does not satisfy the generalized Stokes-Einstein relation at a microscopic scale.
This is a minimal active system, the non-equilibrium  component is included  as active fluctuations that violate fluctuation dissipation relation. Note that the stress in \refn{eq:stress1} should also have a stochastic contribution due to the thermal fluctuations, which we take to be much smaller than the active fluctuations.
	
	Throughout this paper, we work at timescales larger than $\tau_n$ and $\tau_v$, thus~\refn{eq:stress1} becomes 
	\be
	\sigma_{ij} = - p + \eta \l \p_i v_j + \p_j v_i \r + \sigma^n_{ij}.
	\label{eq:stress_fluid1}
	\ee
	with $\Delta(t - t') = 2 \Lambda \delta (t - t')$, obtained by taking the limit 
	$\tau_n \to 0$.
	The dynamics in the Stokes limit is  $\nabla \cdot {\bs \sigma} = 0 $, 
	along with the incompressibility condition $\nabla \cdot {\bs v} = 0$. 
	We emphasize that at long times, the stress as given by \refn{eq:stress_fluid1} is identical to that of a passive viscous fluid, however, as we show in the following, the stochastic dynamics of embedded colloids is, in general, not the same as that of colloids in a passive fluid.

    The generalized Langevin dynamics of a spherical colloidal particle of radius $a$ embedded in an unbounded fluid defined by \refn{eq:stress1} is obtained by integrating out the stress~\refn{eq:stress1} and using no-slip boundary condition at its surface (see, appendix of ref.\,\cite{Lau2009})
    \be 
    \label{eq:gle}
    \tau_m \ddot {\bf R} + \int_{-\infty}^t dt'\gamma(t-t') \dot {\bf R}(t') = - \mu \nabla \cdot {\sf U} + \sqrt{2\lambda \mu} \, {\bs \theta(t)},
    \ee 
    where $\lambda = \Lambda/\eta$, $\mu = 1/6\pi \eta a $, $\tau_m = m\,\mu$, $\gamma(t) =  \tau_v\exp(-t/\tau_v)$ is the friction coefficient, $\mr{\sf U}$ is an externally applied potential, and ${\bs \theta}(t)$ is a Gaussian noise with correlation 
    \be 
    \la \theta_i(t)\theta_j(t')\ra =  \delta_{ij}\tau_n \exp(-(t-t')/\tau_n).
    \ee  
    In the limit $\tau_m,\tau_v, \tau_n \to 0$, \refn{eq:gle} reduces to 
	\be 
	\label{eq:langevin_1p}
	\p_t {\bf R} = - \mu \nabla \cdot {\sf U} + \sqrt{2\lambda \mu} \, {\bs \vartheta},
	\ee  
where ${\bs \vartheta}(t)$ is a unit variance Gaussian white noise.  As expected, the overdamped Langevin dynamics obtained by integrating out the fluid stress~\refn{eq:stress_fluid1} and using no-slip boundary condition at the surface~\cite{Fox1970, Hauge1973, Doi1988} is identical to \refn{eq:langevin_1p}.
Henceforth, wherever required, we use the standard form of Langevin equations corresponding to \refn{eq:stress_fluid1}.

The steady state probability distribution of the position of the colloidal sphere obtained from the  Fokker-Planck equation corresponding to \refn{eq:langevin_1p} has a Boltzmann form $P(z) \propto e^{-\mr{\sf U}/\lambda} $, with an effective temperature $k_B \mr{T_{eff}} \equiv \lambda$.
	\begin{figure}[t!]
		\centering
		\includegraphics[width=\linewidth]{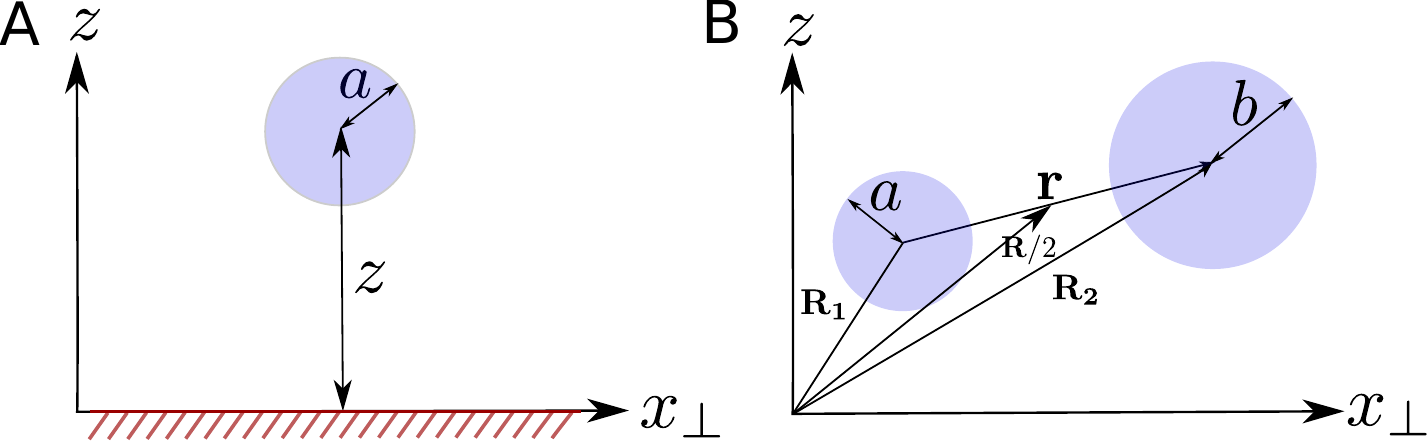}
		\caption{ Schematic of (A) a sphere of radius $a$ at distance $z$ from a fixed wall at $z = 0$, and (B) two spheres of radius $a$ and $b$ centred at ${\bf R_1}$ and ${\bf R_2}$ respectively,  in an unbound fluctuating viscous fluid.
		}
		\label{fig:wall1}
	\end{figure}
    
\section{Particle-Wall Interaction}
	The Langevin dynamics of a spherical colloid of radius $a$ at a distance $z$ from a fixed wall at $z = 0$ (Fig.\,\ref{fig:wall1}(a)),
	is obtained by integrating out  \refn{eq:stress_fluid1} and using the no-slip boundary condition at the surface of the colloid~\cite{Doi1988},
	\bea
	\label{eq:langevin_wall1}
	\p_t  z &=&   - \mr{H}_z \p_z \mr{U}  + \sqrt{2 \lambda \mr{H}_z} \vartheta_z , \\
	\p_t  \xp &=&   - \mr{H}_{\xp} \p_{\xp} \mr{U}  +  \sqrt{2 \lambda \mr{H}_{\xp}}  \vartheta_{\xp}, 
	\label{eq:langevin_wall2}
	\eea
	where $\xp \equiv (x,y)$, $\mr{H}_z (\mr{H}_{\xp})$ is the mobility in the longitudinal (transverse) direction to the wall, $\mr{U}(z,\xp)$ is the 
	particle-wall potential, and
	$\vartheta_z$ and $\vartheta_{\xp}$ are zero mean Gaussian white noise with correlation 
	\be
	\la \vartheta_i(t) \vartheta_j(t') \ra = 2 \lambda \mr{H}^{-1}_{i} \delta_{ij} \delta(t - t'), 
	\label{eq:noise_fluctfluid}
	\ee
	where $i,j \in (x,y,z)$.
	The mobilities $\mr{H}_i(z)$ are functions of the separation $z$ from the wall, which can be evaluated as power series in $a/z$ (Appendix \ref{ap:HI} and~\cite{Kim}).
	The appearance of a multiplicative and correlated noise is typical of a stochastic dynamics with hydrodynamic interactions.  As is well known~\cite{VanKampen1981, Gardiner}, this Langevin equation is meaningless unless supplemented with a stochastic calculus convention for the multiplicative noise.
	The choice of convention depends on the fast timescales that have been integrated out - viscoelastic relaxation time ($\tau_v$), particle inertial relaxation time ($\tau_m \sim m/\eta \, a $, where $m$ is the particle mass and $a$ is the particle size), and noise correlation time ($\tau_n$). 
	In ref.\,~\cite{Rupprecht2018, Vishen2018} it was shown that for an active noise with  $\tau_n \gg \sqrt{\tau_m \tau_v}$, the appropriate 
	convention is Stratonovich~\cite{Stratonovich, Gardiner}. 
	For the cell cortex, for instance, the timescales are $\tau_v \sim1$-$10$\,s~\cite{Saha2016, Rupprecht2018}, $ \tau_m \sim10^{-9}$\,s, and 
	$\tau_n \sim 10$\,s~\cite{Rupprecht2018},
	which makes $\sqrt{\tau_m \tau_v} \sim 0.1\, \mr{ms} \ll \tau_n$.
In the following, we work in the limit $\tau_n \gg \sqrt{\tau_m \tau_v}$, we believe this limit is reasonable for many biological systems.  For the appropriate choice of noise convention when
 this limit does not hold, we refer to \cite{Rupprecht2018, Vishen2018}.

	The Fokker-Planck equation corresponding to~\refn{eq:langevin_wall1} and~\ref{eq:langevin_wall2}, interpreted in Stratonovich convention is~(Appendix \ref{ap:noise_limit})
	\bea
	\nn \p_t P &=& \p_z \l  \mr{H}_z \p_z \mr{U}   + \f{1}{2}\lambda \l \p_z \mr{H}_z \r  +  \lambda \mr{H}_z  \p_z  \r P \\
	&+&  \p_{\xp} \l \mr{H}_{\xp} \p_{\xp} \mr{U}  +  \lambda \mr{ H}_{\xp} \p_{\xp} \r P.
	\label{eq:fp_wall}
	\eea
	which leads to a steady state probability distribution $P(z,x_{\perp}) \propto e^{- \Phi/\lambda}$, with an effective potential,
	\be
	\Phi =  \mr{U} +  \f{1}{2}\lambda \log  \mr{H}_z.
	\label{eq:steadystate1}
	\ee
	This effective interaction  between the wall and the particle, as a result of the active noise and hydrodynamics, is long-range (goes as $a/z$, for large $z$), attractive and anisotropic. 
	The additional fluctuation term along with the applied potential cannot be captured by a simple effective temperature definition. Nevertheless, the steady state has zero current and obeys time reversal symmetry, making this a {\it nonthermal equilibrium model}~\cite{Fodor2016}.
	We emphasize however that unlike in~\cite{Fodor2016}, this deviation from Boltzmann measure occurs {\it even in the limit $\tau_n \to 0$}.
	Further, in contrast to thermal fluctuations where hydrodynamic interactions only affect dynamics, active fluctuations in a fluid affect {\it both} the dynamics and the steady state. 
	
\section{Pressure}	
	The contribution to the force on the wall due to the bare colloid-wall potential $\mr{U}(z)$ is,
	\be
	F_p  =   \int_0^{\infty} dz P(z) \p_z \mr{U}(z) 
	\propto  \int_{0}^{\infty} dz \f{1}{\sqrt{\mr{H}_z}} e^{- \mr{U}/\lambda} \p_z \mr{U}(z) . 
	\ee
	The observation that the force on the wall depends on the form of wall-particle interaction, is directly related to the fact that the probability distribution is non-Boltzmann.
	This is analogous to the situation in dry active particle systems~\cite{Solon2014}, where, apart from the kinetic contribution,
	this would have sufficed to give the pressure.
	However, in momentum conserving Stokesian fluid systems, the net force due to particle-wall interactions is {\it balanced} by the force due to 
	the embedding fluid. 

	Consider a suspension of $N$ colloidal particles in a semi-infinite active fluid confined by a wall.
Assuming particles do not interact directly with each other via forces such as van der Waals, the total force on the $I^{th}$ particle reads:
\be 
\label{eq:totforce_particle}
f_i^{tot} = -\int_{S_p} \sigma_{i j} d S_j + \int_{V_p} d v \, g_i^{p},
\ee
where $\sigma_{ij}$ is the fluid stress acting on the particle $I$, 
$S_p$ is the surface with normal pointing out of the fluid, 
$g^p_1$ is the force density on the colloid due to wall-particle interaction, and 
$V_p$ is the volume of the particle $I$. 
Force balance on the particle $I$: $f_i^{tot} = 0$ implies
\be 
\label{eq:force_particle}
\int_{S_p} \sigma_{i j} d S_j =  \int_{V_p} d v \, g_i^{p}.
\ee 
The dynamics of the Stokesian fluid is given by,
\be 
\label{eq:stress_fluid}
g^f_i + \p_j \sigma_{ij} = 0,
\ee 
where $g^f_i$ is the force density on the fluid due to interaction between fluid particles and the wall. 
Integrating this relation over 
a volume $V$ bounded by two parallel surfaces $S_f$ and $S_f^\prime$, the first at the wall,  the 
second in
the fluid  at a distance such that the body forces $g_i^{p}$ and $g^f_i$ vanish:
\be
\label{eq:force_fluid1}
\int_{S_f} \sigma_{ij} \,  dS_j  +\int_{S_f^\prime} \sigma_{ij} \,  dS_j   + \int_{V - NV_p} dV g^f_i = - \sum_{I = 1}^N \int_{S_p} \sigma_{ij} \, dS_j,
\ee 
where $N$ are the number of colloidal particles within the volume
$V$ and the surface normals point outside the considered fluid volume. 
Substituting \refn{eq:force_particle} in \refn{eq:force_fluid1} gives
\be
\label{eq:force_fluid2}
\int_{S_f} \sigma_{ij} \,  dS  + \int_{S_f^\prime} \sigma_{ij} \,  dS_j= - \int_{V - V_p} dV g^f_i - \sum_{I = 1}^N \int_{V_p} d v \, g_i^{p}.
\ee  
Now, the force exerted by the suspension on the wall is,
\be 
\label{eq:force_wall}
F_{i}=-\int_{S_f} \sigma_{ij} dS_j 
- \int_{V - V_p} dV g^f_i - \sum_{I = 1}^N \int_{V_p} d v \, g_i^{p} ,
\ee 
where the first term is the force on the confining wall due to the fluid, 
the 
second term is the force on the wall due to the fluid particles, which is equal and opposite to the force on the fluid 
due to the wall by Newton third law, similarly, the 
third term is due to the interaction between the colloidal particles and the wall.   
Substituting  \refn{eq:force_fluid2} in \refn{eq:force_wall} we get
\be 
F_{i}= \int_{S_f^\prime} \sigma_{ij} \,  dS_j.
\ee 
Since $S_f^\prime$ is an arbitrary surface chosen to be far from the wall, we see that there is no net added force on the wall even though the particles feel an effective attraction towards it.	The total force exerted on the wall by the system particles$+$fluid vanishes exactly.
This point appears to have been disregarded in recent theoretical studies of the active contribution
	to osmotic pressure and surface tension in momentum conserving active fluids~\cite{Takatori2015, Marconi2015, Rodenburg2017}. 
	
\section{ Two particles embedded in active fluid} 

We now consider the dynamics of two spherical colloids of radius $a$ and $b$, centered at ${\bf R}_1$, and ${\bf R}_2$ respectively (Fig.\,\ref{fig:wall1}(b)),
	in an unbounded active fluctuating fluid.  Since for spherical colloids, the translational motion of the center of mass is decoupled from the rotational motion,  we will consider
	only the hydrodynamic coupling between the translational degree of freedom.
	
	The Langevin dynamics for the centers of the spherical colloids follows as before and is given by~\cite{Kim, Lau2003},
	\be
	\label{eq:langevin_colloids}
	\f{d {\bf R}_{\alpha}}{dt} = \sum_{\beta}{\sf H}_{\alpha \beta} \cdot ( {\bf f}_{\beta} + {\bs \vartheta}_{\beta}) , 
	\ee
	where $\alpha$, and $\beta$ are particle labels, ${\sf H}_{\alpha \beta}$, is  the $3 \times 3$  hydrodynamic interaction tensor coupling the translation motion of particle $\beta$ with that of particle $\alpha$ (see appendix \ref{ap:HI} for the form), ${\bf f_{\beta}} = -\nabla_{{\bf R_{\beta}}} \mr{U}$ is the deterministic force from an externally applied potential $\mr{U}$.
	The stochastic force ${\bs \vartheta}_{\beta}$ on the particle $\beta$, is a three dimensional vector of Gaussian white noise with correlation  
	\be
	\la {\vartheta}_{i\,\alpha}(t) {\vartheta}_{j\,\beta}(s)\ra = 2\lambda {\sf H}_{\alpha\beta}^{-1} \delta_{ij} \delta(t - s). 
	\label{eq:noise_fluctfluid2}
	\ee
	The Fokker-Planck equation corresponding to~\refn{eq:langevin_colloids}, interpreted in  Stratonovich convention is~\cite{Mendez2014} (see appendix \ref{ap:noise_limit})
	\be
	\label{eq:fp_twosphere_1}
	\partial_t P = \nabla_{{\bf R_{\alpha}}} \cdot \ls  - {\sf H}_{\alpha \beta} \cdot {\bf f}_{\beta}  +  \f{\lambda}{2}\l \nabla_{{\bf R_{\beta}}} \cdot  {\sf H}_{\alpha\beta}\r  +  \lambda {\sf H}_{\alpha\beta} \cdot \nabla_{{\bf R_{\beta}}}\rs P. 
	\ee
	Due to incompressibility, $\nabla_{{\bf R_{\beta}}} \cdot {\sf H}_{\alpha\beta}$ is identically zero in Oseen approximation of ${\sf H}_{\alpha\beta}$~\cite{Doi1988}, the first non-zero contribution is at order $1/r^4$. 
	In terms of variables, ${\bf R \equiv R_1 + R_2}$ and ${\bf r \equiv R_2 - R_1}$,~\refn{eq:fp_twosphere_1} is
	\bea
	\label{eq:fp_twosphere_fluid1}
	&&\partial_t P = \nabla_{{\bf R}} \cdot \l - {\sf M}_{11} \cdot {\bf \tilde f}_{1} - {\sf M}_{12} \cdot {\bf \tilde f}_{2}  + \lambda {\sf M}_{11} \cdot \nabla_{{\bf R}} \r P \\ 
	\nn && + \lambda \nabla_{{\bf R}} \cdot \l \f{1}{2} \nabla_{{\bf r}} \cdot {\sf M}_{12}  + {\sf M}_{12} \cdot \nabla_{{\bf r}}  \r P + \nabla_{{\bf r}} \cdot \l \lambda {\sf M}_{12} \cdot \nabla_{{\bf R}} \r P \\
	\nn &&+ \nabla_{{\bf r}} \cdot \l  - {\sf M}_{12} \cdot {\bf \tilde f}_{1} - {\sf M}_{22} \cdot {\bf \tilde f}_{2}  +  \f{\lambda}{2}   \nabla_{{\bf r}} \cdot {\sf M}_{22}  +  \lambda {\sf M}_{22} \cdot \nabla_{{\bf r}}\r P,
	\eea
	where  ${\bf \tilde f}_{1} = -\nabla_{\bf R} \mr{U}$, ${\bf \tilde f}_{2} = -\nabla_{{\bf r}} \mr{U}$, and the mobility matrix
	\bea
	{\sf M} =  \begin{bmatrix}
		{\sf H}_{11} +   {\sf H}_{22}  + 2 {\sf H}_{12}  & {\sf H}_{22} - {\sf H}_{11}  \\
		{\sf H}_{22} - {\sf H}_{11} & {\sf H}_{11} +   {\sf H}_{22}  - 2 {\sf H}_{12}
	\end{bmatrix}.
	\label{eq:HI_matrix1}
	\eea 
	Let us first look at equal sized colloids, $a = b$. In this case, the off-diagonal block matrix ${\sf M_{12}} = 0$, and~\refn{eq:fp_twosphere_fluid1} reduces to 
	\bea
	\label{eq:fp_twosphere_fluid}
	\partial_t P &=& \nabla_{{\bf R}} \cdot \l - {\sf M}_{11} \cdot {\bf \tilde f}_{1}  + \lambda {\sf M}_{11} \cdot \nabla_{{\bf R}} \r P \\ 
	\nn &+& \nabla_{{\bf r}} \cdot \l - {\sf M}_{22} \cdot {\bf \tilde f}_{2}  +  \f{\lambda}{2}   \nabla_{{\bf r}} \cdot {\sf M}_{22}  +  \lambda {\sf M}_{22} \cdot \nabla_{{\bf r}}\r P.
	\eea
	Now, the necessary and sufficient condition for the steady state solution of the Fokker-Plank equation of the from
	\be 
	\p_t P = \sum_{\alpha=1}^{N}\nabla_{\alpha} \cdot \l -{\bf F}_{\alpha} + {\sf D}_{\alpha \beta} \nabla_{\beta}\r P, 
	\ee 
	to have zero probability current is that it satisfies the potential condition~\cite{Gardiner}, defined as,
	\be
	\label{eq:potcondt}
	\f{\p {Z}_{\beta j}}{\p R_{\alpha i}}  =  \f{\p {Z}_{\alpha i}}{\p R_{\beta j}},
	\ee
	where $i,j \in (x,y,z)$ and 
	\be 
	{Z}_{\alpha i} = {\sf D}^{-1}_{\alpha i,  \beta j}{ F}_{\beta j}.
	\ee 
We see that at steady state, \refn{eq:fp_twosphere_fluid} satisfies the  potential condition \refn{eq:potcondt}. 
	Using this, we find that the steady state distribution has the form $P \propto  e^{- \Phi/\lambda}$, 
	where 
	\bea
	\Phi({\bf r, R}) = \mr{U} - \f{1}{2} \lambda \left[ \f{15 b^4}{ 8 r^4} + \mathcal{O}\l \f{1}{r^5}\r \right].
	\label{eq:steadystate_fluid}
	\eea
	As in the colloid-wall interaction, the  steady state distribution is non-Boltzmann
	with a fluctuation-induced  particle-particle interaction that is attractive, though short-ranged. 
		
	For spheres of unequal size, $a \neq b$, the steady state distribution does not obey the potential condition given by \refn{eq:potcondt};
	thus the steady state has a finite probability current and an associated entropy production rate, resulting in a finite propulsion velocity at steady state. We 
	emphasize that unlike AOUP~\cite{Fodor2016}, this nonequilibrium steady state with finite entropy dissipation occurs {\it even in the limit of $\tau_n \to 0$.}
	
	To see this, we describe the two particles as a dimer, characterised by the separation $r$, the orientation ${\bf \hat r}$, and the center of mass position ${\bf R}$~\cite{Illien2017}.  
	In general, it is difficult to obtain an analytic expression for the steady state distribution $P(r, {\bf \hat r}, {\bf R})$. However, in a well defined limit where there is a 
	time scale separation, we obtain analytic expressions for
	the steady state distribution, fluctuation-induced potential and mean propulsion velocity.
	
	Taking $\mr{U}$ to be a function of $r$ alone, ${\bf \tilde f}_1 = 0$ and ${\bf \tilde f}_2 = - \mr{U}'(r) {\bf \hat r}$. With this, the probability flux for {\bf r} in~\refn{eq:fp_twosphere_fluid} is now independent of  {\bf R}, hence, we can integrate out {\bf R} to obtain the marginal dynamics of ${\bf r}$.
	This allows us to solve for the steady state marginal distribution.
	The steady state for the marginal distribution: $P(r)$, obtained by integrating  Eq.\,$14$ in the main text over ${\bf R}$, with no flux boundary condition gives
	\be
	\label{ap:steadyfluid1}
	\nabla_{{\bf r}}  \log P({\bf r}) = \f{1}{\lambda}{\bf \tilde f}_{2} -  \f{1}{2} {\sf M}^{-1}_{22} \cdot \l \nabla_{\bf r} \cdot {\sf M}_{22} \r.
	\ee 
	The mobility tensor ${\sf M}$ defined in Eq.\,$15$  decomposed as sum of projection along $\hat{\bf r}$ (denoted by superscript $n$) and perpendicular to $\hat{\bf r}$ (denoted by superscript $q$) is
	\be
	\label{ap:HI_sum1}
	{\sf M}_{\alpha \beta} = m_{\alpha \beta}^{q}(r) \l {\bf \mathrm{\sf I}} -  {\bf \mathrm{\bf \hat r \hat r}} \r + m_{\alpha \beta} ^{n}(r) {\bf \mathrm{\bf \hat r \hat r}}.
	\ee
	 From this we see that the inverse is
	\be
	{\sf M}_{\alpha \beta}^{-1} = \f{1}{m^q_{\alpha \beta}} \l {\sf I} -  {\bf \hat r \hat r}  \r +  \f{1}{m ^n_{\alpha \beta}}  {\bf \hat r \hat r} , 
	\label{eq:inverse_mobility1}
	\ee
	and its divergence is
	\be 
	\label{eq:divergence_mobility1}
	\nabla_{\bf r} \cdot {\sf M}_{\alpha \beta} = \l \f{\p}{\p r}  m ^n_{\alpha \beta}  +  2 \f{\l  m ^n_{\alpha \beta} -  m^q_{\alpha \beta} \r}{r}   \r {\bf \hat r}.
	\ee 
	
	Using \refn{eq:inverse_mobility1} and \refn{eq:divergence_mobility1} we get 
	\be
	{\sf M}^{-1}_{22} \cdot \l \nabla_{\bf r} \cdot {\sf M}_{22} \r = \f{1}{m_{22} ^n} \l \f{\p}{\p r}  m ^n_{22}  +  2 \f{\l  m ^n_{22} -  m^q_{22} \r}{r}   \r {\bf \hat r}.
	\label{eq:effective_111}
	\ee
	The form of ${\sf M}$ is given in Appendix \ref{ap:HI}. 
 	We see that in Oseen and Rotne-Pragar approximation of ${\sf M}$ the right hand side of \refn{eq:effective_111} is zero. The first nonzero contribution comes when ${\sf M}$ is of order $1/r^4$, at which the self mobilities of the particles are also modified.  Substituting ${\sf M}$ to the order $1/r^4$ (see Appendix \ref{ap:HI}) in \refn{eq:effective_111} gives
	\bea
	{\sf M}^{-1}_{22} \cdot \l \nabla_{\bf r} \cdot {\sf M}_{22} \r  =  \f{ 15 a b (a^3 + b^3)}{2(a + b)} \f{1}{ r^5} \hat {\bf r}.
	\eea
	Substituting this expression in \refn{ap:steadyfluid1} and using ${\bf \tilde f} = - \mr{U}'(r) {\bf \hat r}$ gives
	\be
	\nabla_{{\bf r}}  \log P({\bf r}) = - \f{1}{\lambda} \mr{U}'(r) {\bf \hat r} - \f{ 15 a b (a^3 + b^3)}{4(a + b)} \f{1}{ r^5} \hat {\bf r}.
	\ee 
	Integrating this gives
	\be 
	\label{ap:ss_logprob_fluid}
	\log P({\bf r}) \propto - \f{1}{\lambda} \mr{U}(r) + \f{ 15 a b (a^3 + b^3)}{16(a + b)} \f{1}{ r^4},
	\ee 
	which gives $P({r}) \propto e^{-\Phi({\bf r})/\lambda}$, where 
	\be
	\Phi = \mr{U}( r) -  \f{1}{2} \lambda \left[ \f{ 15 a b (a^3 + b^3)}{8(a + b)} \f{1}{ r^4}  + \mathcal{O}\l \f{1}{r^5}\r \right].
	\label{eq:steadystate_fluid1}
	\ee
	Note that while we can define an effective potential for the marginal dynamics of $r$ , there is no   
	effective potential description in the full Fokker-Planck description that includes 
	$\hat {\bf r}$ and ${\bf R}$. If we 
	now assume that the dynamics of $r$ is {\it fast}, 
	we can decompose the probability distribution as 
		\be
	P({\bf R, \hat r}, r) = P(r) \int dr r^2 P({\bf R, \hat r}, r) = P(r) \tilde P({\bf R, \hat r}), 
	\ee 
	where we have defined $ \tilde P({\bf R, \hat r}) \equiv \int dr r^2 P({\bf R, \hat r},r) $.
	
	Decomposing the derivative as radial and rotational derivative 
	\be
	\nabla_{{\bf r}} = {\bf \hat r} \f{\p}{\p r} + \f{1}{r}{\bf\hat r} \times \mathcal{R},  
	\ee
	where $\mathcal{R}$ is the rotation operator that rotates the ${\bf r}$ keeping $r$ fixed. In spherical co-ordinates it reads
	\be 
	{\bf\hat r} \times \mathcal{R} = \hat \theta \f{\p}{\p \theta} +  \hat \phi \f{1}{\sin \theta}\f{\p}{\p \phi}.
	\ee 
	Integrating out $r$ from \refn{eq:fp_twosphere_fluid} we get \cite{Illien2017}
	\bea
	\label{ap:fp_fluid_int}
	\nn\partial_t \tilde P &=& \nabla_{{\bf R}} \cdot \ls  - v_0 \, {\bf \hat r}  + \lambda\lla \f{{\sf M}_{12}}{r}  \rla \cdot {\bf \hat r}\times \mathcal{R}  + \lambda \lla {\sf M}_{11} \rla \cdot \nabla_{{\bf R}} \rs \tilde P \\ 
	&+&  \lambda {\bf \hat r}\times \mathcal{R}  \cdot \ls \lla \f{{\sf M}_{12}}{r}   \rla \cdot \nabla_{{\bf R}} + \lla \f{{\sf M}_{22}}{r^2}  \rla \cdot {\bf \hat r}\times \mathcal{R} \rs \tilde P , 
	\eea
	where the averages are over the distributions $P(r)$, $\la \phi(r) \ra = \int dr r^2 \phi(r) P(r)$, and the self-propelled velocity
	\be
	\label{eq:ssp_fluid}
	v_0 = \int r^2 dr {\bf \hat r} \cdot \l {\sf M}_{12} \cdot {\bf \tilde f}_{2} -  \f{\lambda}{2} \nabla_{{\bf r}} \cdot {\sf M}_{12} - \lambda \sf{M}_{12} \cdot \nabla_{{\bf r}} \r P(r).
	\ee
	Note that if the fluctuations are thermal, the middle term in \refn{eq:ssp_fluid} is not present, 
	the probability distribution has the Boltzmann form $P(r) \propto e^{- \mr{U}/\lambda}$, and the velocity vanishes as shown Appendix \ref{ap:force_fluid}.
	To evaluate the average, we consider the bare inter-particle potential to be a stiff spring, $\mr{U} = k (r - l)^2/2$, and $l \gg a,b$. In this limit, the relaxation time 
	scale of $r$ is set by $\sf{M_{22}}$, $k$, and $\lambda$; $k l^2/2\lambda  \gg 1$ ensures that the $r$-dynamics is fast~\cite{Illien2017}. 
	Using the Laplace approximation~\cite{Bender2013}, we obtain, to leading order in $1/l$, the propulsion velocity 
	\be
	v_0 =  \f{5 a b}{4 \pi \eta} (b - a) \f{ \lambda}{ l^5}  + \mathcal{O}\l \f{1}{l^6}\r.
	\ee
	directed along ${\bf \hat r}$, if $b > a$.
	Orientation decorrelation will lead to diffusion over times longer than the orientation correlation  time of ${\bf \hat r}$ ($\tau_R$). The enhancement of the diffusion constant due to this self-propulsion in units of diffusion of a sphere of radius $l$ is $v_0^2 \tau_R/D_t = (1 - \delta )^2 \delta^2 \epsilon^6 $, where $D_t = \lambda/6\pi \eta l$, $\tau_R \sim 3\eta l^3/\lambda $, $\delta = a/b$, and $\epsilon = b/l$. Since $\delta < 1$ and $\epsilon \ll 1$,
	this enhancement is  very small.
	
	\section{ Activating the particle by a fluctuating force} 
	Now consider two colloidal spheres of radii $a$ and $b$, embedded in an unbounded passive fluid, each of which
	experiences a stochastic force, {\it localized on the colloids}. 
	To make the discussion simple, we set the stress fluctuations of the embedding medium $\sigma^n_{ij} = 0$; 
	the dynamics of the colloids is then given by Eq.\,\ref{eq:langevin_colloids}, with ${\bs \vartheta_{\beta}} = 0$, and the applied force on the particles is a sum of deterministic and stochastic components,
	${\bf f_{\beta}} \equiv {\bf f_{\beta}} + {\bs \xi_{\beta}}$. The fluctuating force on the two colloids is taken to be isotropic, zero mean Gaussian white with correlations,
	\be
	\la {\bf \xi}_{\alpha i}(t) {\bf \xi}_{\beta j }(s)\ra = 2 \Lambda \delta_{\alpha \beta} \delta_{ij}\delta(t - s).
	\label{eq:noise_fluctforce2}
	\ee
	Since the fluctuation is external and the dissipation is in the fluid bath, the system is always active. Hence, for any choice of noise convention the fluctuation dissipation relation is not satisfied. For consistency with the approximation for passive fluid, we take the noise correlation time to be the slowest timescale leading to Stratonovich noise convention.
	The corresponding Fokker-Planck equation (in Stratonovich convention) is now (see Appendix \ref{ap:noise_limit}),
	\bea
	\label{eq:fp_fluct_force}
	&&\partial_t P= \nabla_{{\bf R}} \cdot \l - {\sf M}_{11} \cdot {\bf \tilde f}_{1} - {\sf M}_{12} \cdot {\bf \tilde f}_{2}  + \Lambda {\sf D}_{11} \cdot \nabla_{{\bf R}} \r P \\ 
	\nn && + \Lambda \nabla_{{\bf R}} \cdot \l   \f{1}{2} \nabla_{{\bf r}} \cdot {\sf D}_{12}  +  {\sf D}_{12}  \cdot \nabla_{{\bf r}} \r P + \nabla_{{\bf r}} \cdot \l \Lambda {\sf D}_{12} \cdot \nabla_{{\bf R}} \r P \\
	\nn &&+ \nabla_{{\bf r}} \cdot \l  - {\sf M}_{12} \cdot {\bf \tilde f}_{1} - {\sf M}_{22} \cdot {\bf \tilde f}_{2}  + \f{1}{2}  \Lambda \nabla_{{\bf r}} \cdot {\sf D}_{22}  +  \Lambda {\sf D}_{22} \cdot \nabla_{{\bf r}}   \r P,
	\eea
	where, ${\sf M}$ is given by Eq. \ref{eq:HI_matrix1}, and the diffusion matrix ${\sf D}_{\alpha \beta}$ is $3\times3$ matrix given in terms of ${\sf H}_{\alpha \beta}$ by ${\sf D}_{11} = \l {\sf H}_{11} + {\sf H}_{12} \r^2 + \l {\sf H}_{22} +  {\sf H}_{12} \r^2$, ${\sf D}_{12} = {\sf D}_{21} = {\sf H}_{22}^2  -  {\sf H}_{11}^2$, and ${\sf D}_{22} = \l {\sf H}_{11} - {\sf H}_{12} \r^2 + \l {\sf H}_{22} -  {\sf H}_{12} \r^2$.
	
	Once again, Eq.\,\ref{eq:fp_fluct_force} does not satisfy the potential condition given by \refn{eq:potcondt}, and hence does not have a zero probability current steady state,
	{\it even when the spheres are of the same size}. This proves that the dynamics of particles in an active medium is fundamentally
	different from the dynamics of activated particles.
In this context, we refer to recent experiments~\cite{ Berut2014, Berut2016} in which two spheres embedded in a fluid are held in two optical traps. A fluctuating force is applied on one sphere by 
	moving the position of its laser trap randomly. This has been modeled as a two temperature system~\cite{Hough2002,Berut2014, Berut2016}, where the static particle feels the bath temperature and the particle in the fluctuating trap, a higher temperature.
	Our study demonstrates the inadequacy of such an effective temperature approach, and in principle (at least numerically) provides a full solution to the steady state 
	distribution.
	
	We now consider the case when $\mr{U}$ is a function of $r$ alone implying ${\bf \tilde f}_1 = 0$ and ${\bf \tilde f}_2 = - \mr{U} {\bf \hat r}$. For this case \refn{eq:fp_fluct_force} reduces to 
\bea
\label{eq:fp_fluct_force2}
\nn \partial_t P &&= \nabla_{{\bf R}} \cdot \l  - {\sf M}_{12} \cdot {\bf \tilde f}_{2} + \f{1}{2} \nabla_{{\bf r}} \cdot {\sf D}_{12} + \Lambda {\sf D}_{11} \cdot \nabla_{{\bf R}} \r P \\ 
\nn && + \Lambda \nabla_{{\bf R}} \cdot \l  {\sf D}_{12}  \cdot \nabla_{{\bf r}} \r P + \nabla_{{\bf r}} \cdot \l \Lambda {\sf D}_{12} \cdot \nabla_{{\bf R}} \r P \\
&&+ \nabla_{{\bf r}} \cdot \l  - {\sf M}_{22} \cdot {\bf \tilde f}_{2}  + \f{1}{2}  \Lambda \nabla_{{\bf r}} \cdot {\sf D}_{22}  +  \Lambda {\sf D}_{22} \cdot \nabla_{{\bf r}}   \r P.\qquad
\eea
The steady state for the marginal distribution $P(r)$ obtained by integrating  \refn{eq:fp_fluct_force2} over ${\bf R}$, with no flux boundary condition gives
\be 
\label{ap:ss_fluct_force}
\nabla_{{\bf r}} \log P({\bf r}) = \f{1}{\Lambda} {\sf D}_{22}^{-1}\cdot {\sf M}_{22} \cdot {\bf \tilde f}_{2} - \f{1}{2} {\sf D}_{22}^{-1} \cdot \nabla_{{\bf r}} \cdot {\sf D}_{22}.
\ee 
The tensor ${\sf D}$ decomposed as sum of projection along $\hat{\bf r}$ (denoted by superscript $n$) and perpendicular to $\hat{\bf r}$ (denoted by superscript $q$) is
\be
\label{eq:Dtensor}
{\sf D}_{\alpha \beta} = d_{\alpha \beta}^{q}(r) \l {\bf \mathrm{\sf I}} -  {\bf \mathrm{\bf \hat r \hat r}} \r + d_{\alpha \beta} ^{n}(r) {\bf \mathrm{\bf \hat r \hat r}}.
\ee
The inverse is
\be
{\sf D}_{\alpha \beta}^{-1} = \f{1}{d^q_{\alpha \beta}} \l {\sf I} -  {\bf \hat r \hat r}  \r +  \f{1}{d ^n_{\alpha \beta}}  {\bf \hat r \hat r} , 
\label{eq:inverse_diffusion}
\ee
and the divergence is
\be
\label{eq:divergence_diffusion}
\nabla_{\bf r} \cdot {\sf D}_{\alpha \beta} =  \l \f{\p}{\p r}  d ^n_{\alpha \beta}  +  2 \f{\l  d^n_{\alpha \beta} -  d^q_{\alpha \beta} \r}{r}   \r {\bf \hat r}.
\ee
Using Eq. \ref{eq:inverse_diffusion}, and Eq. \ref{eq:divergence_diffusion} we get 
\be
\label{ap:fluct_effect1}
{\sf D}^{-1}_{22} \cdot \l \nabla_{\bf r} \cdot {\sf D}_{22} \r = \f{1}{d_{22} ^n} \l \f{\p}{\p r}  d ^n_{22}  +  2 \f{\l  d ^n_{22} -  d^q_{22} \r}{r}   \r {\bf \hat r}.
\ee
Taking the diffusion tensor $\sf{D}$ (see Appendix \ref{ap:force_particle}) to the order $1/r$ and substituting it in \refn{ap:fluct_effect1} gives
\be
\label{ap:fluct_effect2}
{\sf D}^{-1}_{22} \cdot \l \nabla_{\bf r} \cdot {\sf D}_{22} \r  = -\f{9(a^2 b^2)}{4(a^2 + b^2)} \f{1}{r^3} \hat {\bf r},
\ee
and substituting the diffusion tensor in the first term on the right of \refn{ap:ss_fluct_force} we obtain
\be 
\label{ap:fluct_det}
{\sf D}_{22}^{-1}\cdot {\sf M}_{22} \cdot {\bf \tilde f}_{2} = - \f{m_{22}^n}{d_{22}^n} {\sf U'}(r) \hat {\bf r}
\ee
Substituting \refn{ap:fluct_effect2} and \ref{ap:fluct_det} in \refn{ap:ss_fluct_force} gives
\be
\label{ap:ss_fluct_force1}
\nabla_{{\bf r}}  \log P({\bf r}) = - \f{1}{\Lambda}\f{m_{22}^n}{d_{22}^n} {\sf U'}(r) {\bf \hat r} + \f{9(a^2 b^2)}{8(a^2 + b^2)} \f{1}{r^3} \hat {\bf r}.
\ee 
From this the effective potential $\Phi \equiv - \Lambda \log P$  upon integration of \refn{ap:ss_fluct_force1} is 
		\be
\label{eq:ss_pot_force}
\Phi = \int dr r^2 \f{m_{22}^n}{d_{22}^n}{\mr U'} + \Lambda\f{9(a^2 b^2)}{16(a^2 + b^2)} \f{1}{r^2} + \mathcal{O}\l \f{1}{r^3}\r.
\ee 
Note that in this case $m_{22}/d_{22}$ depends on $r$, it is not possible to define an effective free energy, keeping the energy $\mr{U}$ and a constant effective temperature.
As in the colloid-colloid interaction in an active fluid, the  steady state distribution is non-Boltzmann
with a fluctuation-induced  particle-particle interaction. However, in contrast, the interaction is repulsive, long-ranged, and depends on the form of the interaction potential $\mr{U}$.
This effect is similar to that of effective colloid-wall interactions due interplay between hydrodynamic and electrostatic interactions \cite{Saha2014}.

Integrating out $r$ from \refn{eq:fp_fluct_force2} leads to the similar form of self-propulsion velocity as  \refn{eq:ssp_fluid}. 
To order $1/r^3$ ${\sf M}_{12}$ is a constant and self-propulsion velocity is
\be
v_0 = - N \int dr r^2 \l m^n_{12} \mr{U}' -  d^n_{12}  \f{\p}{\p r} \Phi \r  e^{-\Phi/\Lambda}.
\ee
Using \refn{eq:ss_pot_force} we get
\be
\f{\p}{\p r} \Phi = \f{m_{22}^n}{d_{22}^n}{\mr U'} - \Lambda\f{9(a^2 b^2)}{8(a^2 + b^2)} \f{1}{r^3} + \mathcal{O}\l \f{1}{r^4}\r.
\ee 
and
\be 
\f{m_{22}^n}{d_{22}^n} =  \f{(6 \pi \eta b a)(a + b)}{a^2 + b^2}  \left[  1 +  \f{6 b^2 a^2}{(a^2 + b^2)(a + b)}\f{1}{r}  + \mathcal{O}\l \f{1}{r^2}\r \right].
\ee 
Thus to leading order in $1/r$ we obtain
\be
v_0 = - N \int dr r^2 m^n_{12} \l 1 - \f{ d^n_{12}  m^n_{22}}{ d^n_{22}  m^n_{12}}\r{\mr U'}  e^{- {\mr U}/\Lambda'},
\ee
where $\Lambda' = \Lambda (a^2 + b^2)/6 \pi \eta  a b(a + b)$.
As before, taking
$\mr{U} = k (r - l)^2/2$, with $k$ large and $l \gg a,b$, such that $k l^2/2\Lambda  \gg 1$, gives the mean self-propulsion velocity of the dimer to be
		\be
		\label{eq:sp_ezyme}
		\nn v_0 \sim \f{a - b}{18 \pi^2 \eta^2  a b(a + b)}  \f{\Lambda}{l},
		\ee
		directed along $\hat {\bf r}$. Note that this leading order contribution is due to the interaction potential between the dimer. In contrast, the leading order contribution in the fluctuation fluid case was from the fluctuation induced interaction.
	The long time dynamics of the dimer is diffusive, thus resembling an active Brownian particle~\cite{Romanczuk2012}.
	The enhancement in diffusion constant over the bare diffusion
	 $D_t \sim \mr{k_B T}/6\pi \eta l$ is  
	 \be 
	 \f{v_0^2 \tau_R}{D_t} \sim  \f{(1 - \delta)^2}{(1 + \delta^2)^2 \epsilon^2} \f{\Lambda'^2}{\mr{k_B T}^2}
	 \ee 
	where the rotational correlation time is $\tau_R \sim 3\eta l^3/\lambda $, $\delta = a/b$, $\epsilon = b/l$, and $\Lambda' = \Lambda (a^2 + b^2)/6 \pi \eta  a b(a + b)$. Since both $\mr{k_B T}/\Lambda', \epsilon \ll 1$, this enhancement can now be large compared to the bare diffusion.	
	For $\Lambda' \sim \mr{k_B T}$, $a = 2 \, \mr{nm}$, $b = 3 \, \mr{nm}$, $l = 10 \, \mr{nm}$, and $\eta = 10^{-3} \,\mr{Pa\,s}$ we get $v_0 = 5 \, \mr{mm/s}$ and $v_0^2 \tau_R \sim 25 \, \mu\mr{m^2/s}$, a value which is comparable to thermal diffusivity $D_t$. 

\subsection{One-Dimensional example of self-propulsion}
To obtain an intuitive understanding of the self-propulsion described above,  consider a simple example in which two spherical colloids interacting via a harmonic potential are confined along the $x$-axis. The center of the spheres are positioned at $x_1(t)$ and $x_2(t)$ with $x_1 < x_2$. This dimer is activated by an external force $f$ acting on the colloid at $x_1$. 
The one-dimensional Langevin dynamics for this dimer, with the hydrodynamic interaction tensor \cite{Kim} (Appendix \ref{ap:HI}) can be obtained from \refn{eq:langevin_colloids},
\bea 
\dot x_1 &=& \l\f{1}{6\pi \eta a} - \f{1}{4\pi\eta r }\r k (r-l) + \f{f}{6\pi \eta a}, \\ 
\dot x_2 &=& -\l\f{1}{6\pi \eta a} - \f{1}{4\pi\eta r }\r k (r-l) + \f{f}{4\pi \eta a},
\eea 
where $r = x_2 - x_1$. In terms of the separation $r$ and center of mass $R = x_1 + x_2$ we get 
\bea 
\dot r &=& -\l\f{1}{6\pi \eta a} - \f{1}{4\pi\eta r }\r (2 k (r-l) + f), \\ 
\dot R &=& \l\f{1}{6\pi \eta a} + \f{1}{4\pi\eta r }\r f.
\eea 
At steady state $r = l - f/2k$ and the center of mass velocity is
\be 
\dot R =  \f{f}{6\pi\eta a} + \f{f}{4\pi \eta \l l - f/2k \r}.
\ee 
Thus we see that the magnitude of $\dot R$ are not equal for $f \to -f $. For positive $f$ the harmonic spring is compressed leading to a larger speed in comparison to the case when $f$ is negative and the spring is stretched leading to a lower speed. 
Now consider a periodic symmetric driving where $f$ switches direction after fixed time intervals of duration $\Delta t$. The average force applied is zero, however, the dimer will have a net positive velocity given by
\be 
\la \dot R \ra =  \f{f^2/k}{8\pi \eta \l l^2 - (f/2k)^2 \r}.
\ee 
This mechanism is similar to that proposed in ref.\,\cite{Kumar2008,Baule2009} for the self-propulsion of an elastic dimer on a frictional substrate. 
This setup can be experimentally realized by connecting a paramagnetic and a diamagnetic bead by a polymer and applying a magnetic field gradient which periodically switches direction. 

\section{Discussion} 
	
	We have seen that the interplay between
	nonequilibrium fluctuations and hydrodynamics, even in the simplest extension of a single particle embedded in a momentum conserving fluid,
	viz., two particles embedded in an unbounded active gel or a particle in an active gel bound within confining walls, 
	brings out the inadequacy of the effective temperature description, since the corresponding steady state probability distribution shows strong departures from the equilibrium Boltzmann form. 
   This leads to a non-equilibrium effective ``Casimir-like''  power law interaction 
   ~\cite{Kirkpatrick2013, Aminov2015, Cattuto2006, Bartolo2003, Parra-Rojas2014, Brito2007}. 
   	 Furthermore, both in an active fluid or in a passive fluid activated by stochastic forcing, unequal size particles exhibit short time ballistic motion. 
This shows that the effective equilibrium limits, that have been obtained for ``dry" active systems, do not hold for ``wet" active systems.
	The effective interaction between activated particles can be experimentally verified by more precise measurements of the particle position in the optical trap setups used in the experiments reported in Ref.\, \cite{Berut2016}.
	
	We expect the interplay between hydrodynamic interactions and active fluctuations will also lead to effective interaction between self-propelled particles in a passive fluid. Since the effective interactions are sensitive to the precise origin of the fluctuation, the form of effective interaction may be different for different models of self-propelled particles. 
	There are various extensions of this work that will be useful to explore.  
	So far in this work we only look at incompressible fluid,  it would be of interest to analyze the behavior for compressible fluids. 
	The form of the effective interaction and self-propulsion velocity was computed in the far-field approximation of the hydrodynamic interaction. We expect the effective interaction and self-propulsion to exist in lubrication limit as well. 

\section{Acknowledgements} 
We thank M. Wyart, J.-F. Rupprecht, R. Morris, K.B. Husain and  S.A. Rautu for useful discussions. 
	
\appendix
	\section{Fokker-Planck from Langevin} \label{ap:noise_limit}
	Following \cite{Mendez2014} we derive the Fokker-Planck equation from a multivariate overdamped Langevin equation with multiplicative noise, for general choice of stochastic calculus.
	The Langevin equations are 
	\bea
	\dot x_i = F_i({\bf x}) + G_{ij}({\bf x})\, \vartheta_j,
	\label{eq:multiplicative}
	\eea
	where $i \in (1,..., N)$ and $j \in (1,...,M)$, $\vartheta_j$ is a zero mean Gaussian white noise with correlation 
	\be 
	\la \vartheta_k(t) \vartheta_j(t') \ra = C_{kj}({\bf x}) \delta(t - t'),
	\ee 
	where $(k,j) \in (1,...,M)$.
	Integrating \refn{eq:multiplicative} over a small time interval $\Delta t$ gives
	\be
	\Delta x_i = \int_{t}^{t + \Delta t} dt' F_i({\bf x}) + \int_{t}^{t + \Delta t} dt'\, G_{ij}({\bf x}) \vartheta_j.
	\ee
	The first term on the right is unambiguously approximated for small $\Delta t$ using a straightforward Taylor expansion of $F_i$, in contrast, the  limit of the second term is not well defined \cite{Gardiner, Lau2007}. 
	Unlike deterministic calculus, there are multiple choices for stochastic calculus.  
	This amounts to choosing the time between $t$ and $t + \Delta t$ at which ${\bf x}$ in $G({\bf x})$ is evaluated.
	If $G({\bf x})$ is evaluated at $t$ it is  Ito calculus \cite{Gardiner}, at $t + \Delta t/2$ it is Stratonovich calculus \cite{Stratonovich,Gardiner}, and at $t + \Delta t$ it is Hanggi-Klimontovich calculus \cite{Hanggi1982,Klimontovich1990}. Using the definition introduced in \cite{Lau2007}, ${\bf x}(t)$ evaluated at any generic point between $t$ and $t + \Delta t$, parameterized by $\epsilon \in (0,1)$ gives 
	\be
	\Delta x_i =  F_i({\bf x}_t) \Delta t +  G_{ij}({\bf x}_t + \epsilon \Delta {\bf x}))\int_{t}^{t + \Delta t} dt' \vartheta_j(t').
	\ee
	Taylor expanding $G({\bf x})$ around ${\bf x}(t)$
	\bea
	\nn \Delta x_i &=& F_i({\bf x}_t) \Delta t +  G_{ij}({\bf x}_t)\int_{t}^{t + \Delta t} dt' \vartheta_j(t') \\
	&+& \epsilon \f{\p G_{ij}({\bf x}_t)}{\p x_k} \Delta x_k\int_{t}^{t + \Delta t} dt' \vartheta_j(t').
	\eea
	$\Delta x_k$ has a term of order $\sqrt{\Delta t}$ hence the $\epsilon$ term has a contribution of order $\Delta t$, substituting $\Delta x_k$ back in the equation and keeping terms to order $\Delta t$ we get
	\bea
	 \Delta x_i &=&  F_i({\bf x}_t) \Delta t +  G_{ij}({\bf x}_t)\int_{t}^{t + \Delta t} dt' \vartheta_j(t') \\
	\nn &+& \epsilon \f{\p G_{ij}({\bf x}_t)}{\p x_k} G_{kl}({\bf x}_t)\int_{t}^{t + \Delta t} dt' \vartheta_l(t') \int_{t}^{t + \Delta t} dt'' \vartheta_j(t'') .
	\eea
	The first and the second moment  of $\Delta x$ are 
	\bea
	&&\la \Delta x_i \ra  =  F_i({\bf x}_t) \Delta t  +  \epsilon \f{\p G_{ij}({\bf x}_t)}{\p x_k} G_{kl}({\bf x}_t) C_{lj}({\bf x}_t) \Delta t ,\quad \\
	&&\la \Delta x_i \Delta x_l \ra = G_{ij} G_{lk} C_{jk} \Delta t.
	\eea
	The corresponding Fokker-Plank equation is \cite{Gardiner}
	\be
	\partial_t P = \frac{\partial}{\partial x_i} \l - F_i  - \epsilon \f{\partial G_{ik}}{\partial x_j} S_{jk}+ \frac{1}{2} \frac{\partial}{\partial x_j} G_{ik}S_{jk} \r P , 
	\label{eq:FP_general}
	\ee
	where $S_{ij} = C_{ik} G_{jk}$.
	Thus we see that different choices of stochastic calculus ($\epsilon$) leads to different Fokker-Planck equations and hence different physics. 
	
	For a given problem, the relevant value of $\epsilon$ depends on the fast timescales which have been integrated out. Furthermore the existence of a simple convention choice is not always guaranteed \cite{Rupprecht2018}.  In this paper, the effective description of \textcolor{green}{a}viscoelastic gel is obtained by integrating out: the viscoelastic relaxation time ($\tau_v$), the inertial relaxation time ($\tau_m = m/\eta$), and the noise correlation time ($\tau_n$). 
	In \cite{Rupprecht2018, Vishen2018} it was shown that for an exponentially correlated noise with $\tau_n \gg \sqrt{\tau_m \tau_v}$ $\epsilon = 1/2$ (Stratonovich convention) is the right value.
	For  thermal noise, in general, no simple convention 
	works. But if $\f{G_{ik}}{\p x_j}S_{jk} = G_{ik}\f{S_{jk}}{\p x_j}$, $\epsilon = 1$ (Hanggi-Klimontovich convention) is the right value (see \cite{Lau2007} for a detailed discussion on noise convention for thermal fluctuations).
	These are the convention choices used throughout this paper. 
	
	The Langevin equations in the main text are of the form given by \refn{eq:multiplicative} and the corresponding Fokker-Planck will be given by \refn{eq:FP_general} by making the following identifications in different cases

	\begin{itemize}
		\item {\bf Wall-particle}:
		Comparing \refn{eq:multiplicative} with \refn{eq:langevin_wall1},\ref{eq:langevin_wall2}  and noise correlation given by \refn{eq:noise_fluctfluid} we get: ${\bf F} =  - \nabla \mr{U} \, \hat z$, ${\sf G} = {\sf H} $, and ${\sf C} = 2 \lambda {\sf H}^{-1}$, which gives ${ \sf S} = 2 \lambda {\sf H}^{-1}\cdot {\sf H} = 2\lambda \, {\sf I}$,where ${\sf I}$ is $3\times 3$ identity matrix.  Using these values in \refn{eq:FP_general} leads to  Eq.\,$8$.
		
		\item {\bf Fluctuating fluid}:
		Comparing \refn{eq:multiplicative} with \refn{eq:langevin_colloids} and noise correlation given by \refn{eq:noise_fluctfluid2}  we get:  ${\sf G} = {\sf H} $, and ${\sf C} = 2 \lambda {\sf H}^{-1}$, and ${\sf S} = {\sf C \cdot H}  = 2\lambda {\sf I} $, where ${\sf I}$ is $6\times 6$ identity matrix.
		Using the above values in \refn{eq:FP_general} we get Eq.\,$13$.
		
		\item  {\bf Fluctuating force}:
		Comparing \refn{eq:multiplicative} and Eq.\,$11$ and the noise correlation given by \refn{eq:noise_fluctforce2}  we get:  ${\sf G} = {\sf H} $, and ${\sf C} = 2 \Lambda {\sf I}$, and ${\sf S} = {\sf C \cdot H}  = 2\Lambda {\sf H} $, where ${\sf I}$ is the $6\times 6$ identity matrix.
		Using the above values in \refn{eq:FP_general} and changing of variables to ${\bf R}$ and ${\bf r }$ we get \refn{eq:fp_fluct_force}.
	\end{itemize}

	
	\section{Hydrodynamic interaction tensor}\label{ap:HI}
	
	\subsection{Wall-Particle}
	
	For distances larger than the particle size ($z \gg a$), the mobilities can be calculated as a power series in the inverse of separation from the wall ($1/z$). To second order (Rotne-Prager approximation) the mobility longitudinal to the wall is \cite{Kim,Rotne1969} 
	\be
	\mr{H}_{z}  = \f{1}{6 \pi \eta a} \l 1 - \f{9}{8}\f{a}{z} + \f{1}{2}\f{a^3}{z^3}\r,
	\ee
	and the mobility transverse to the wall is
	\be
	{\mr H}_{x_{\perp}} = \f{1}{6 \pi \eta a} \l 1 - \f{9}{16}\f{a}{z} + \f{1}{8}\f{a^3}{z^3}\r.
	\ee

	\subsection{Two spheres}

	The hydrodynamic interaction tensor ${\sf H_{\alpha \beta}}$ coupling the translational degree of freedom is a $3 \times 3$ tensor. 
	This tensor can be calculated as a power series in the inverse of separation ($1/r$) between the center of the two spheres.  To the fourth power it is given by \cite{Kim} 
	\bea
	\label{eq:HI_order4}
	\nn {\sf H_{11}}  &=&  \f{1}{6 \pi \eta a } ({\sf I} - {\bf \hat r \hat r})  +  \l \f{1}{6 \pi \eta a }  - \f{5}{8}\f{b^3}{\pi \eta r^4} \r {\bf \hat r \hat r},\\
	\nn {\sf H_{22}} &=&  \f{1}{6 \pi \eta b } ( {\sf I} - {\bf \hat r \hat r} ) +   \l \f{1}{6 \pi \eta b }  - \f{5}{8}\f{a^3}{\pi \eta r^4} \r {\bf \hat r \hat r},
	\eea
	\bea
	\nn {\sf H_{12}} = {\sf H_{21}} &=&  \l \f{1}{8 \pi \eta r } +  \f{1}{24} \f{(a^2 + b^2)}{\pi \eta r^3} \r ({\sf  I} - {\bf \hat r \hat r} ) \\
	&+&   \l \f{1}{4 \pi \eta r } +  \f{1}{12} \f{(a^2 + b^2)}{\pi \eta r^3}\r {\bf \hat r \hat r}.
	\eea
	
	The mobility tensor ${\sf M}$ defined in Eq.\,$15$ of the main text decomposed as sum of projection along $\hat{\bf r}$ (denoted by superscript $n$) and perpendicular to $\hat{\bf r}$ (denoted by superscript $q$) is
	\be
	\label{ap:HI_sum}
	{\sf M}_{\alpha \beta} = m_{\alpha \beta}^{q}(r) \l {\bf \mathrm{\sf I}} -  {\bf \mathrm{\bf \hat r \hat r}} \r + m_{\alpha \beta} ^{n}(r) {\bf \mathrm{\bf \hat r \hat r}}.
	\ee
	By substituting \refn{eq:HI_order4} into \refn{ap:HI_sum} we get
	\bea
	\label{eq:mob1}
	m_{11}^q  &=&     \f{(a + b)}{6 \pi \eta a b} + \f{1}{4 \pi \eta r } + \f{1}{12} \f{(a^2 + b^2)}{\pi \eta r^3} ,\\
	m_{11} ^n  &=&    \f{(a + b)}{6 \pi \eta a b} + \f{1}{2 \pi \eta r }  - \f{1}{6} \f{(a^2 + b^2)}{\pi \eta r^3}  -  \f{5}{8}\f{a^3 + b^3}{\pi \eta r^4},  \qquad \\
	m_{12}^n &=&  \f{a - b}{6 \pi \eta a b } + \f{5}{8}\f{(b^3 - a^3)}{\pi \eta r^4} , \\
	\label{eq:mob_m12}
	m_{12}^q &=&  \f{a - b}{6 \pi \eta a b }, \\
	m_{22}^q  &=&     \f{(a + b)}{6 \pi \eta a b} - \f{1}{4 \pi \eta r } - \f{1}{12} \f{(a^2 + b^2)}{\pi \eta r^3} ,\\
	m_{22} ^n  &=&    \f{(a + b)}{6 \pi \eta a b} - \f{1}{2 \pi \eta r }  + \f{1}{6} \f{(a^2 + b^2)}{\pi \eta r^3} -  \f{5}{8}\f{a^3 + b^3}{\pi \eta r^4} . \qquad
	\label{eq:mob2}
	\eea

	\section{Passive colloids in active fluid - Thermal Fluctuations}\label{ap:force_fluid}

	For thermal fluctuations, $\lambda = k_B T$, and the Fokker-Planck equation corresponding to \refn{eq:langevin_colloids} with noise correlation given by \refn{eq:noise_fluctfluid2} in Hanggi-Klimontovich convention is
	\bea
	\label{ap:fp_thermal1}
	\nn \partial_t P &&= \nabla_{{\bf R}} \cdot \l  - {\sf M}_{12} \cdot {\bf \tilde f}_{2} +\lambda  {\sf M}_{12} \cdot \nabla_{{\bf r}} + \lambda {\sf M}_{11} \cdot \nabla_{{\bf R}} \r P \\ 
	&&+ \nabla_{{\bf r}} \cdot \l  - {\sf M}_{22} \cdot {\bf \tilde f}_{2}  + \lambda {\sf M}_{12} \cdot \nabla_{{\bf R}} +  \lambda {\sf M}_{22}  \cdot  \nabla_{{\bf r}} \r P. \qquad
	\eea
	The marginal of steady state distribution of $r$ is $P(r) \propto e^{- \mr{U}/\lambda}$.
	Integrating out separation $r$ as for active fluctuation we get the self-propulsion velocity of the form
	\be
	v_0 = \int r^2 dr {\bf \hat r} \cdot \l {\sf M}_{12} \cdot {\bf \tilde f}_{2} - \lambda \sf{M}_{12} \cdot \nabla_{{\bf r}} \r P(r).
	\ee
	Expanding this we get
	\be
	v_0 = -\int dr r^2 \l m^n_{12} \mr{U}' e^{- \mr{U}/\lambda}  +  \lambda  m_{12}^n  \f{\p}{\p r} e^{- \mr{U}/\lambda} \r = 0.
	\ee
	As expected for thermal fluctuations the self-propulsion velocity is identically zero.

	\section{Activated particles in a passive fluid}\label{ap:force_particle}
	
	The effective diffusion tensor for an activated particle in a passive fluid as defined in \refn{eq:fp_fluct_force} is 
	\be
	{\sf D} =  \begin{bmatrix}
		{\sf L}_{11} +  {\sf L}_{22}  +  {\sf L}_{12} +   {\sf L}_{21}   & {\sf L}_{22}  - {\sf L}_{11}  \\
		{\sf L}_{22}  - {\sf L}_{11}  & {\sf L}_{11}  +   {\sf L}_{22}   -  {\sf L}_{12} -  {\sf L}_{21} 
	\end{bmatrix}, 
	\label{eq:diffusion_matrix_2}
	\ee
	where ${\sf L}$ in terms of the hydrodynamic interaction tensor ${\sf H}$ is
	\be
	{\sf L} =   \begin{bmatrix}
		{\sf H}_{11}\cdot {\sf H}_{11} +  {\sf H}_{12} \cdot {\sf H}_{21}  & {\sf H}_{11}\cdot {\sf H}_{12} +  {\sf H}_{12} \cdot {\sf H}_{22} \\
		{\sf H}_{21}\cdot {\sf H}_{11} +  {\sf H}_{22} \cdot {\sf H}_{21}  &  {\sf H}_{12}\cdot {\sf H}_{21} +  {\sf H}_{22} \cdot {\sf H}_{22}
	\end{bmatrix}. 
	\ee
	
	Substituting in \refn{eq:diffusion_matrix_2} ${\sf H}$ from \refn{eq:HI_order4} and keeping terms only to order $1/r$  gives
	\bea
	\nn {\sf D}_{11} &=&  \l \l  \f{1}{6 \pi \eta a } + \f{1}{8 \pi \eta r } \r^2 +  \l  \f{1}{6 \pi \eta b } + \f{1}{8 \pi \eta r } \r^2 \r \l {\sf I} - {\bf \hat r \hat r}  \r  \,\, \\
	&+&  \l \l  \f{1}{6 \pi \eta a } + \f{1}{4 \pi \eta r } \r^2 +  \l  \f{1}{6 \pi \eta b } + \f{1}{4 \pi \eta r } \r^2 \r {\bf \hat r \hat r}, \\
	\label{ap:diff_11}
	\nn {\sf D}_{12} &=& {\sf D}_{21} =  \l \f{1}{(6 \pi \eta b)^2 } - \f{1}{(6 \pi \eta a)^2 } \r \l {\sf I} - {\bf \hat r \hat r} \r \\
	&+&  \l \f{1}{(6 \pi \eta b)^2 } - \f{1}{(6 \pi \eta a)^2 } \r  {\bf \hat r \hat r}, \\
	\label{ap:diff_12}
	\nn {\sf D}_{22} &=&  \l \l  \f{1}{6 \pi \eta a } - \f{1}{8 \pi \eta r } \r^2 +  \l  \f{1}{6 \pi \eta b } - \f{1}{8 \pi \eta r } \r^2 \r \l {\sf I} - {\bf \hat r \hat r}  \r \\
	&+&  \l \l  \f{1}{6 \pi \eta a } - \f{1}{4 \pi \eta r } \r^2 +  \l  \f{1}{6 \pi \eta b } - \f{1}{4 \pi \eta r } \r^2 \r {\bf \hat r \hat r} .
	\label{ap:diff_22}
	\eea

%


\end{document}